\documentclass[showkeys,showpacs,longbibliography,prb]{revtex4-2}
\usepackage{amsmath}
\usepackage{xcolor}
\usepackage{hyperref}
\usepackage{graphicx}
\usepackage{ulem}

\newcommand{\VEC}[1]{{\boldsymbol{ #1}}}

\newcommand{\ie}{{\it i.e.}}
\newcommand{\vs}{{\it vs.}}
\newcommand{\Fig}{{Fig.}}
\newcommand{\Eq}{{Eq.}}
\newcommand{\Eqs}{{Eqs.}}
\newcommand{\be}{\begin{equation}} \newcommand{\ee}{\end{equation}}
\newcommand{\bea}{\begin{eqnarray}} \newcommand{\eea}{\end{eqnarray}}
\newcommand{\gru}{Gr\"uneisen}
\newcommand{\this}{GQA}
\newcommand{\toher}{QDM}

\begin{document}
\title{
A size-consistent \gru{}-quasiharmonic approach for lattice thermal conductivity
}

\author{Chee Kwan Gan}
\email{ganck@ihpc.a-star.edu.sg}
\affiliation{Institute of High Performance Computing, 1 Fusionopolis Way, \#16-16 Connexis 138632, Singapore}
\date{Nov 8, 2022}

\author{Eng Kang Koh}
\affiliation{School of Physical and Mathematical Sciences, Nanyang Technological University, 21 Nanyang Link 637371, Singapore}

\begin{abstract}
We propose a size-consistent \gru{}-quasiharmonic approach (\this{}) to calculate the lattice thermal conductivity $\kappa_l$
where the \gru{} parameters that measure the degree of phonon anharmonicity
are calculated directly using first-principles calculations.
This is achieved by identifying and modifying two existing equations related to the Slack formulae for $\kappa_l$
that suffer from the size-inconsistency problem  when dealing with non-monoatomic 
primitive cells (where the number of atoms in the primitive cell $n$ is greater than one).
In conjunction with other thermal parameters such as the acoustic Debye temperature $\theta_a$ that can also be obtained within the \this{}, we 
predict $\kappa_l$ 
for a range of materials taken from the diamond, zincblende, rocksalt, and wurtzite compounds.
The results are compared with that from the experiment and the quasiharmonic Debye model (\toher{}).
We find that in general the prediction of $\theta_a$ is
rather consistent among the \this{},  experiment, and \toher{}.
However, while the \toher{} somewhat overestimates the \gru{} parameters 
and hence underestimates $\kappa_l$ for most materials,
the \this{} predicts the experimental trends of \gru{} parameters
and $\kappa_l$ more closely. 
We expect the \this{} with the modified Slack formulae could be used as an effective and practical
predictor for $\kappa_l$, especially for crystals with large $n$.

\end{abstract}


\maketitle
\section{Introduction}
Thermal conductivity of a material is of fundamental interest and crucial for the
performance of thermal management, heat dissipation,
thermoelectrics, 
and thermal barrier coatings\cite{Yan22v21,Kundu21v126,Chen19v170,Miller17v29,Yu16v6,Waldrop16v530,Nath17v129,Tadano18v120,Seko15v115}.
Determining lattice thermal conductivity $\kappa_l$ from a truly first-principles approach
requires handling of phonon-phonon interactions\cite{Lindsay16v20,Ward09v80} that
is extremely compute intensive and therefore cannot be done in a routine manner.
To treat many systems in a reliable and computationally efficient manner a few feasible methods have been proposed.
These include the quasiharmonic Debye model (QDM)\cite{Toher14v90}, the relaxation-time approximation (RTA) method\cite{Bjerg14v89}, 
and a modified Debye-Callaway model\cite{Gorai17v2}.
Among these the QDM is interesting due to its high computing efficiency. The workhorse of
the QDM for calculating $\kappa_l$ is the Slack formulae\cite{Morelli06} that require
two key input parameters, \ie, the acoustic Debye temperature $\theta_a$ and the \gru{} parameters.
These two quantities could be deduced approximately
by an equation-of-state analysis\cite{Blanco04v57}
which requires the total energy as a function of volume that may be easily obtained 
from only first-principles total energy calculations, thereby avoiding the moderately expensive phonon calculations.
Even though the QDM somewhat consistently underestimates $\kappa_l$ due to
overestimations of \gru{} parameters, it can still reliably predict the ordinal ranking of
$\kappa_l$ for several classes of semiconductor materials. It was shown\cite{Toher14v90} that the QDM is able to
deliver high Pearson correlation between the predicted $\kappa_l$ and experimental $\kappa_l$.
In view of the advantages of the \toher{}, it is interesting to ask if
the prediction of $\kappa_l$ could be further improved when
$\theta_a$ and the \gru{} parameters are
obtained from the standard first-principles phonon calculations.
One may argue that phonon calculations may be still be expensive for a routine calculation, but they are
considered to be very inexpensive compared to a full density-functional theory (DFT)
anharmonic treatment\cite{Garg11v106} that accounts for multiple-phonon scatterings. Indeed,
phonon databases\cite{MatProj-link,Petretto18v5,TogoPhononDB2020-link} have been
established that show the feasibility of standard phonon calculation route
with today's computing resources.
We note that systematic \gru{} parameter determinations have also been feasibly carried out
for the evaluation of the thermal expansion coefficients 
within the \gru{} formalism.\cite{Schelling03v68,Ding15v5,Gan15v92,Liu17v121,Gan18v151,Liu18v154,Gan19v31,Liu19v99,Gan22v12}

The outline of this paper is as follows. In Section~\ref{sec:method} we 
review the existing Slack formulae in estimating $\kappa_l$.
The size consistency issue of the original Slack formulae is then 
diagnosed and discussed.  Subsequently we propose the necessary modifications, 
using the specific heat argument, to the existing Slack formulae to preserve size consistency.
In Section~\ref{sec:results} we present the results obtained using the proposed \this{}.
Section~\ref{sec:summary} contains conclusions and a summary.

\section{Methodology}
\label{sec:method}

The lattice thermal conductivity $\kappa_l(T)$ at temperature $T$ according to the Slack formula~\cite{Morelli06} is given by
\be
\kappa_l(T) = \kappa_l(\theta_a) \frac{\theta_a}{T}
\label{eq:kappal_T}
\ee
Here $\theta_a$ is the acoustic Debye temperature\cite{Morelli06} that can be deduced from the Debye temperature $\theta_D$ and 
the number of atoms in a primitive cell $n$ according to
\be
\theta_a = \frac{\theta_D}{n^{\frac{1}{3}}}
\label{eq:theta_a}
\ee
We note there are different phononic properties, one 
related to the heating and another to heat conduction, that are appropriately described by, respectively,
the Debye temperature $\theta_D$, and the acoustic Debye temperature $\theta_a$ where the optical modes
contribute little to the heat conduction.
At $T= \theta_a$, $k_l(\theta_a)$ is given by\cite{Leibfried54v71,Julian65v137,Slack79v34,Morelli06}
\bea
\kappa_l(\theta_a) &=&
\left( \frac{ 0.849 }{ 1 - 0.514\gamma_a^{-1} + 0.228 \gamma_a^{-2}   } \right)
\frac{3 \times 4^{\frac{1}{3}} }{ 20 \pi^3 }
\frac{ k_B^3 }{ \hbar^3  } 
\frac{\theta_a^2 }{\gamma_a^2 }\frac{M}{n} V^{\frac{1}{3}}
\label{eq:kappal_at_theta_a}
\eea
where $\gamma_a$  is the macroscopic \gru{} parameter $\gamma_m(T)$ evaluated at $ T = \theta_a$ 
(we shall use \Eq~\ref{eq:gp_mac} to evaluate $\gamma_m(T)$). The primitive 
cell has a mass $M$ and a volume $V$. $ h = 2\pi \hbar $ and $k_B$ are the
Planck and Boltzmann constants, respectively.

In this work, we obtain $\theta_D = h\nu_{c}^{\rm min}/k_B $ by finding the Debye cutoff frequency $\nu_{c}^{\rm min}$ that 
gives the best match\cite{Gan18v151} between 
the constant-volume heat capacity $C_v^D(\nu_c,T)$ from the Debye model
and 
the constant-volume heat capacity $C_v^{DFT}(T)$  from the DFT.
This is achieved through minimizing the absolute error
\be
d(\nu_c) = \frac{1}{(3n k_B)^2} \int_0^\infty dT [ C_v^D (\nu_c, T) - C_v^{DFT}(T)]^2
\ee
According to the Debye model\cite{McQuarrie2000-book}, the Debye 
phonon density of states $\rho_D(\nu) = A\nu^2$ for $0 \le \nu \le \nu_c$ where $A = 9n/\nu_c^3$ to ensure normalization, 
and $\rho_D(\nu) $ vanishes outside the $0 \le \nu \le \nu_c$ range. This allows the calculation of
\be
C_v^D(\nu_c,T) = \int_0^{\nu_c} \rho_D(\nu) c(\nu,T) d\nu
\ee
Here the heat capacity contributed by a phonon mode of frequency $\nu$ is $c(\nu,T) = k_B r^2/\sinh^2 r$, $r=h\nu/2k_B T$. 
From the DFT calculations, we obtain
\be
C_v^{DFT}(T)  =  \frac{V}{(2\pi)^3} \sum_\lambda \int_{\rm BZ} d\VEC{q}\   c(\nu_{\lambda\VEC{q}},T) 
\ee
The summation is over all phonon mode indices $\lambda$ and wavevector $\VEC{q}$ in the Brillouin zone (BZ). 
We note that $C_v^{DFT}(T)$ may also be determined using the DFT phonon density of states information $\rho_{DFT}(\nu)$ where
\be
C_v^{DFT}(T)  =  \int_0^{\nu_{\rm max}} \rho_{DFT}(\nu) c(\nu,T) d\nu
\ee
where $\nu_{\rm max}$ is the maximum frequency in the phonon spectrum.

The mode \gru{} parameter is defined as $\gamma_{\lambda\VEC{q}} = - (V/\nu_{\lambda\VEC{q}}) \partial \nu_{\lambda\VEC{q}}/\partial V$. 
With $\gamma_{\lambda\VEC{q}}$, the macroscopic \gru{} parameter\cite{Gan18v151,AshcroftMermin1976-book} as a function of $T$ is
\be
\gamma_{m} (T) = 
\frac{ \sum_\lambda \int_{\rm BZ} d\VEC{q}\  \gamma_{\lambda\VEC{q}} c(\nu_{\lambda\VEC{q}},T)    }
{ \sum_\lambda \int_{\rm BZ} d\VEC{q}\   c(\nu_{\lambda\VEC{q}},T)    }
\label{eq:gp_mac}
\ee

Now we discuss the size-consistency issue.
\Eq~\ref{eq:kappal_at_theta_a} was originally derived for the monoatomic case, where $n=1$. 
While the efficacy of the Slack formulae (\Eqs~\ref{eq:kappal_T}, \ref{eq:theta_a}, 
and \ref{eq:kappal_at_theta_a}) have been well established\cite{Morelli06} when $n=2$, we note that 
\Eq~\ref{eq:theta_a} may not be appropriate for cells with large $n$ since $\theta_D$ depends on the atomic mass
and bond strength but is independent\cite{Morelli06} of $n$. In fact we expect the expression for 
$\theta_a$ is to be independent of $n$ if
we assume there are two crystals that are similar physically but differing only in $n$. 
Take the rocksalt NaCl crystal as an example. We could choose a primitive cell with two atoms ($n=2$) to do a first calculation.
However, we could also use the conventional cubic cell with eight atoms ($n=8$)
to do a second calculation and we 
expect $\theta_a$ to be the same as that from the first calculation. 
In Section~~\ref{sec:results} we will 
show the heat capacities at $T=\theta_a$ deduced from \Eq~\ref{eq:theta_a}
for some wurtzite compounds with $n=4$ is 
indeed smaller compared with that of the
zincblende and rocksalt compounds with $n=2$.
Next, because of \Eqs~\ref{eq:theta_a} and ~\ref{eq:kappal_at_theta_a},  $\kappa_l(\theta_a)$ 
is seen to scale as $n^{-\frac{1}{3}}$, which is undesirable since 
we would expect $\kappa_l(\theta_a)$ to be also independent of $n$.
To overcome this size-consistency issue
while retaining the good efficacy\cite{Morelli06} of the method when
$n=2$, we propose to define
\be
\theta_a' = \frac{\theta_D}{\Lambda}
\label{eq:modified_theta_a}
\ee
with $\Lambda = 2^{\frac{1}{3}}$.
This choice of $\Lambda$ is made so that we can retain the 
same numerical results as that obtained with the original Slack formulae in order to
make a direct comparison between the
\this{} (that uses the DFT input parameters) and \toher{} for the case of $n=2$. The optimal choice of $\Lambda$ may be 
further investigated in a future study.
To be consistent with \Eq~\ref{eq:modified_theta_a}
we restore the size-consistency of \Eq~\ref{eq:kappal_at_theta_a} by using the following modified equation
\bea
\kappa_l'(\theta_a') &=&
\left( \frac{ 0.849 }{ 1 - 0.514(\gamma_a')^{-1} + 0.228 (\gamma_a')^{-2}   } \right)
\frac{3 \times 4^{\frac{1}{3}} }{ 20 \pi^3 }
\frac{ k_B^3 }{ \hbar^3  } 
\frac{\theta_a'^2 }{(\gamma_a')^2 }\frac{M}{n^{\frac{4}{3}}} V^{\frac{1}{3}} 2^{\frac{1}{3}}
\label{eq:kappal_modified}
\eea
where $\gamma_a' = \gamma_m(\theta_a')$. Notice that the exponent of $n$ in the above equation has been changed from $1$ in \Eq~\ref{eq:kappal_at_theta_a} to $\frac{4}{3}$ to achieve size consistency.
The size consistency of \Eq~\ref{eq:kappal_modified} is seen by observing that both $M$ and $V$ scale linearly with $n$, therefore
the appearance of $MV^{1/3}$ in the numerator cancels out the effect of $n^{\frac{4}{3}}$ in the denominator.
With this modification, we obtain a modified equation for the lattice thermal conductivity where
\be
\kappa_l'(T) = \kappa_l'(\theta_a') \frac{\theta_a'}{T}
\ee

In this work, since the calculations for each compound involve many steps,
it is important to design a workflow that will automatically generate the structures required for atomic 
relaxations and phonon calculations. It also needs to 
handle automatic job submissions and post-processing of DFT results such as scanning of Debye cutoff frequencies 
to find the Debye temperature.

\section{Results}
\label{sec:results}
\begin{figure}
\centering\includegraphics[width=9.2cm,clip]{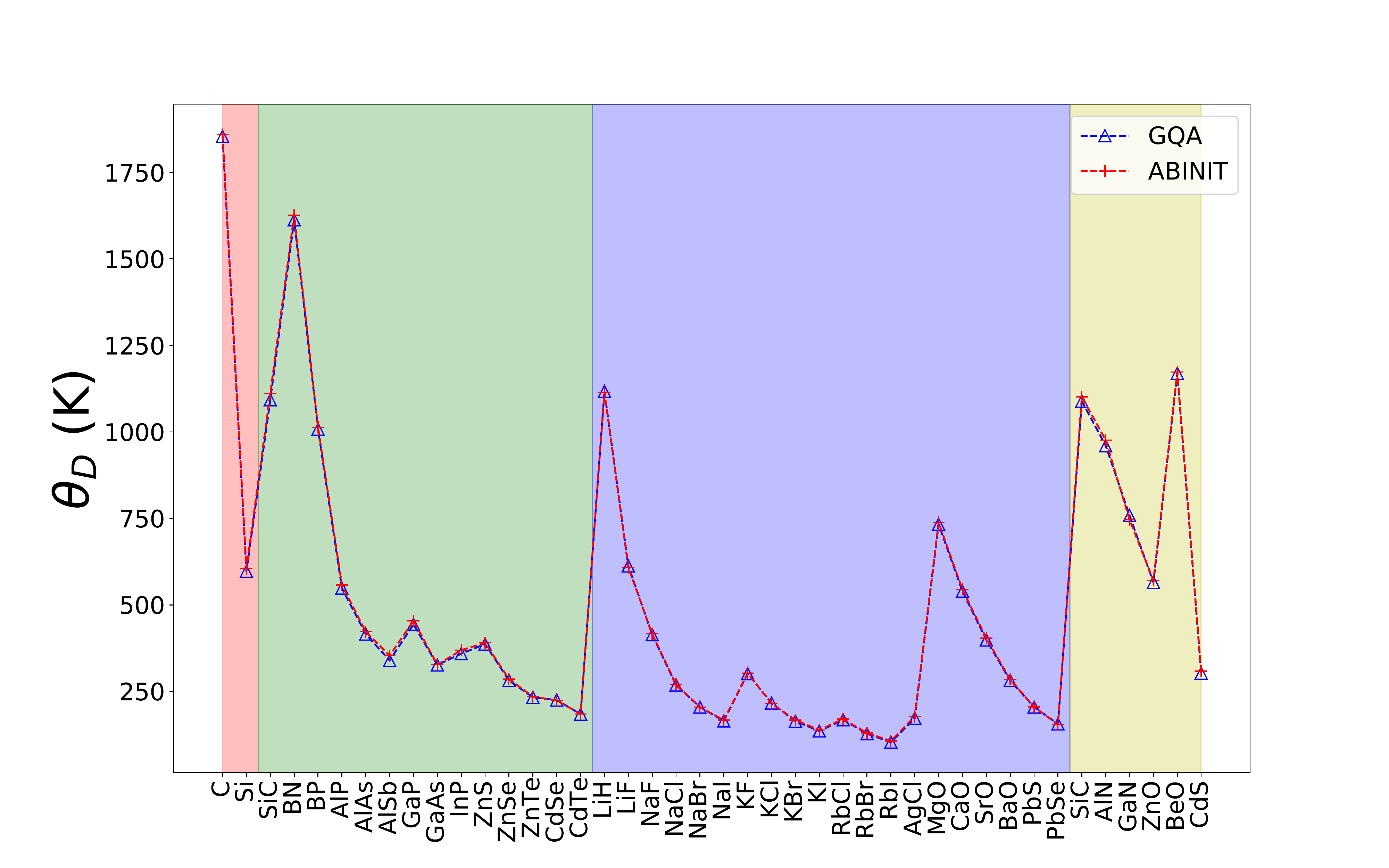}
\caption{The Debye temperature $\theta_D$ determined from the phonon density of states 
for the VASP (\this{}) and  ABINIT calculations\cite{Petretto18v5}.
Each compound is grouped according to the class it belongs. From the left to right, the sequence is diamond, zincblende, rocksalt, and wurtzite.
}
\label{fig:001-TD.pdf} 
\end{figure}
We use the optimized structures from a phonon database~\cite{Petretto18v5} as a starting point
to perform the necessary atomic and cell relaxations.
Density-functional theory (DFT) calculations are carried out
as implemented in the Vienna Ab initio Simulation
Package (VASP)\cite{Kresse96v6} with projected augmented-wave pseudopotentials
and a PBEsol\cite{Perdew08v100} functional.
A relatively high cutoff energy of $600$~eV is used throughout. 
Atomic relaxation is stopped when the maximum force on all atoms
is less than $10^{-3}$~eV/\AA{}.
Phonon calculations are performed with a small-displacement method\cite{Kresse95v32,Gan21v259}.
We have used a supercell of $2\times 2 \times 2$ for the cubic based compounds (\ie, diamond, zincblende,
and rocksalt) and $ 3 \times 3 \times 2$ for the wurtzite compounds. 

From the phonon density of states obtained from the DFT calculations we 
carry out the error
minimization as outlined in Section~\ref{sec:method} to find
the Debye frequency $\nu_c^{\rm min}$ 
and then determine the Debye temperature $\theta_D = h \nu_c^{\rm min}/k_B$.
Note that most numerical values related to the \this{} have been tabulated in the Supplementary Material.
\Fig~\ref{fig:001-TD.pdf},
shows the Debye temperatures $\theta_D$ obtained from the 
VASP and the ABINIT codes that are very similar 
even though one is based on the small-displacement method (VASP) and the other
on the density-functional perturbation
theory (ABINIT)\cite{Petretto18v5}. As expected, the compounds with large $\theta_D$ tend to consist of only very light elements.

\begin{figure}
\centering\includegraphics[width=9.2cm,clip]{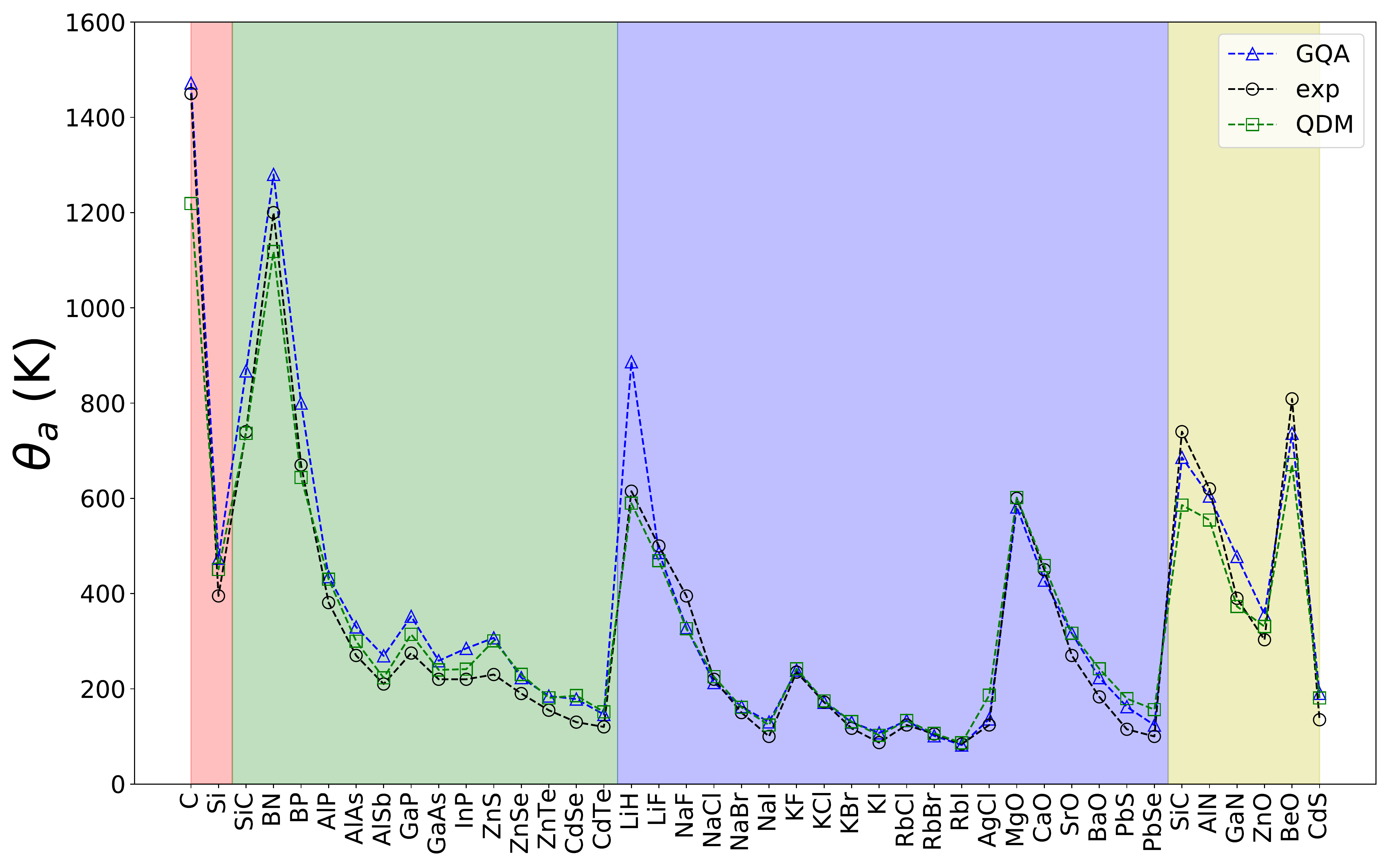}
\caption{The acoustic Debye temperature $\theta_a$ for compounds from the \this{} (this work), experiment (exp), and 
the \toher{}.
Each compound is grouped according to the class it belongs. From the left to right, the sequence is diamond, zincblende, rocksalt, and wurtzite.
}
\label{fig:T-acous-by-TD-div-n-raised-third.pdf}
\end{figure}

\Fig~\ref{fig:T-acous-by-TD-div-n-raised-third.pdf} shows the results of $\theta_a$ from
the \this{} and \toher{}\cite{Toher14v90} (that are deduced from Eq~\ref{eq:theta_a}), as well as that from the experiment.
We note that the estimation of $\theta_a$ from the \toher{} is rather impressive 
even though it uses only the energy \vs{} volume information (\ie, without any direct 
phonon information). We find that the \this{} delivers $\theta_a$ that is somewhat larger than that from experiment for the zincblende compounds. However,
the agreement is better for the rocksalt compounds. 
The agreement between $\theta_a$ 
from the \this{} and experiment is poor for compounds involving very light elements (such as LiH),
presumably due to the presence of very high-frequency phonons.
We note that better estimates of $\theta_a$ may be achieved by a proper modification of Eq~\ref{eq:theta_a} for each crystal class. 
However, since we aim to report the results obtained from the original Slack formulae, 
we leave the improvement of $\theta_a$ estimation in a future study. 

\begin{figure}
\centering\includegraphics[width=9.2cm,clip]{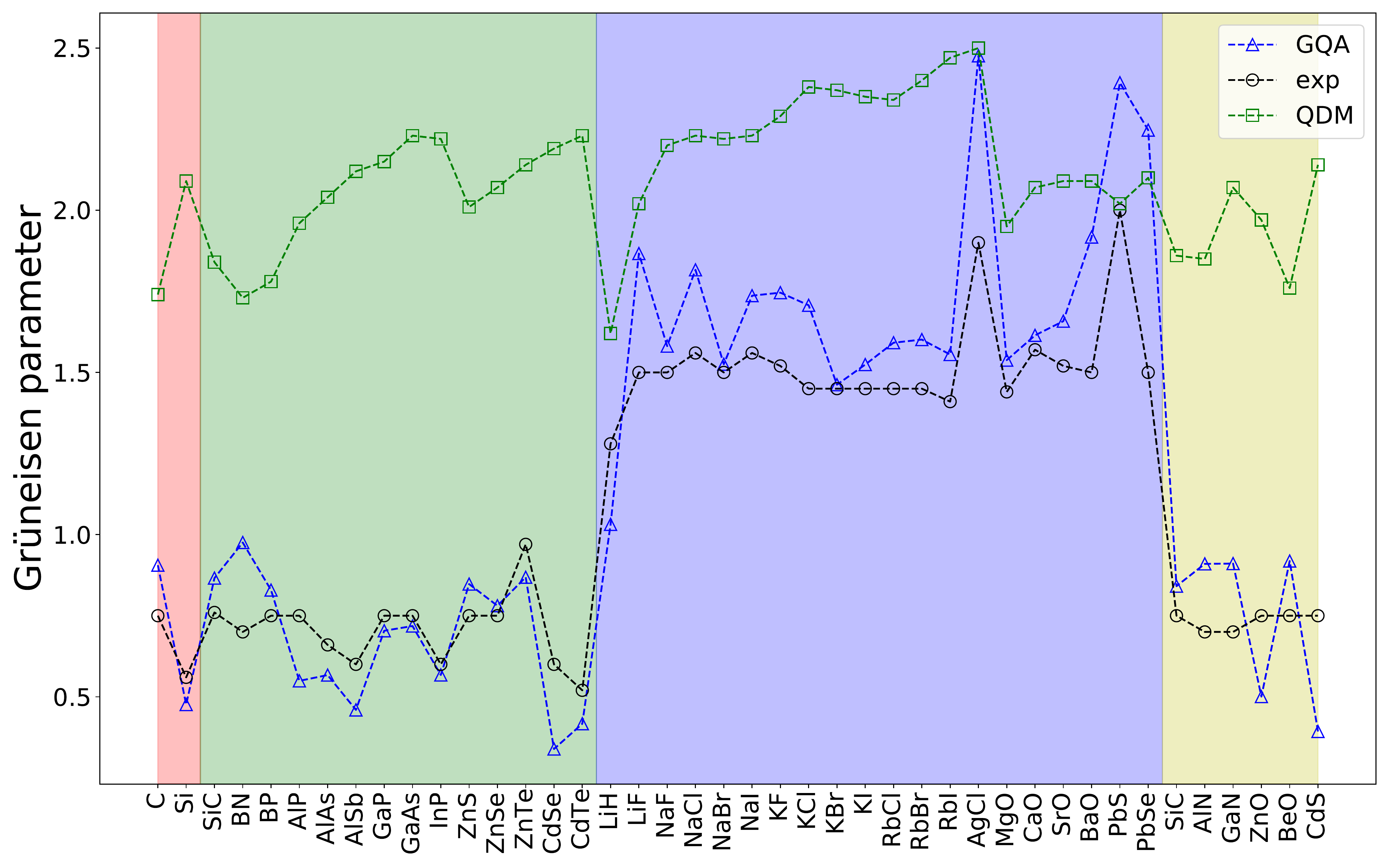}
\caption{\gru{} parameters of compounds from the \this{}, experiment (exp), and \toher{}.
Each compound is grouped according to the class it belongs. From the left to right, the sequence is diamond, zincblende, rocksalt, and wurtzite.
}
\label{fig:GP-by-TD-div-n-raised-third.pdf}
\end{figure} 

\begin{figure}
\centering\includegraphics[width=9.2cm,clip]{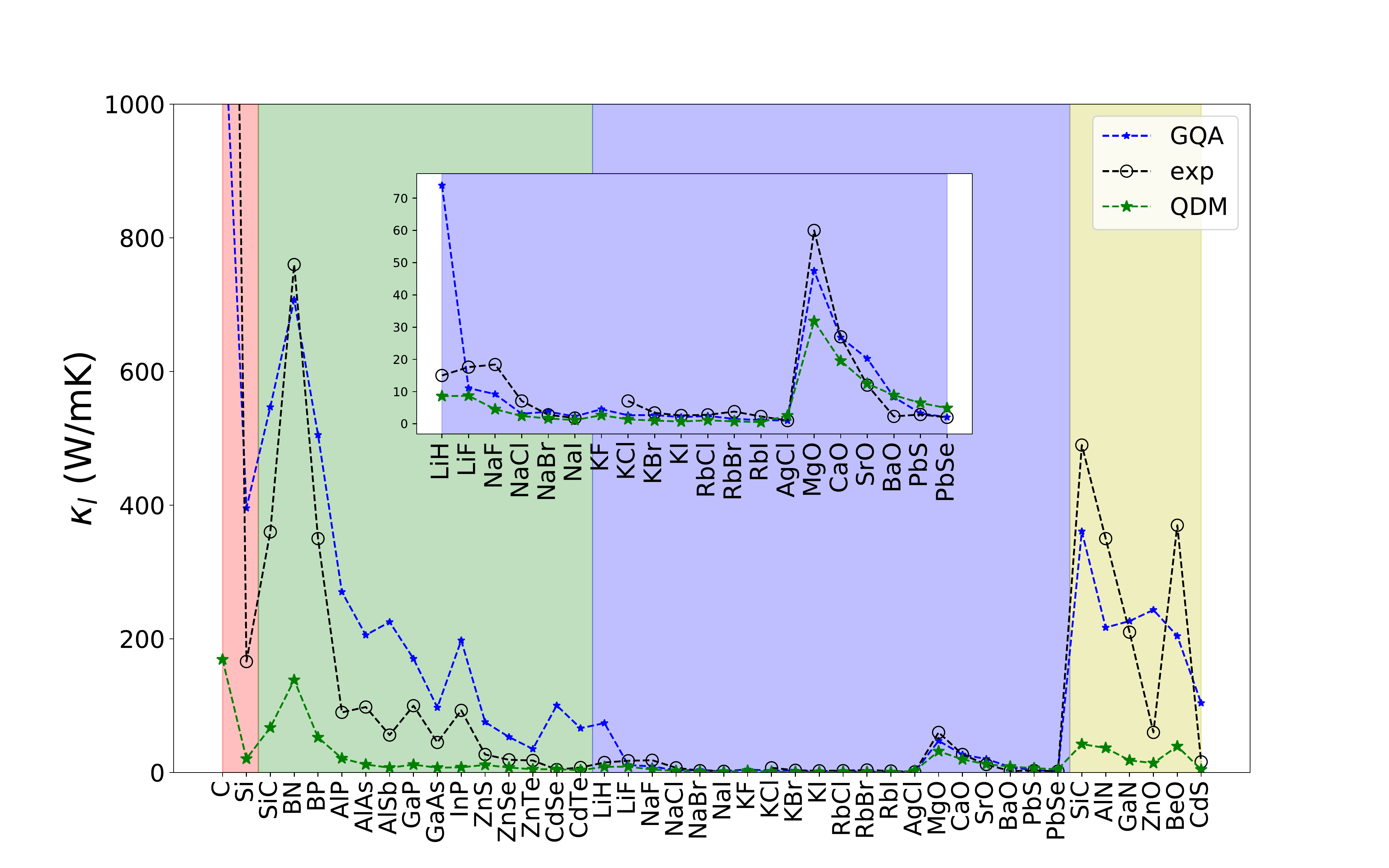}
\caption{The lattice thermal conductivity $\kappa_l$ at $T=300$~K 
from experiment and theoretical predictions. The inset shows $\kappa_l$ of the rocksalt compounds.
Each compound is grouped according to the class it belongs. From the left to right, the sequence is diamond, zincblende, rocksalt, and wurtzite.
}
\label{fig:LTC-by-TD-div-n-raised-third.pdf}
\end{figure}

The \gru{} parameters are calculated using a finite central-difference scheme 
with strains of $\pm 1\%$ and a phonon-band connectivity technique.\cite{Gan21v5}.
\Fig~\ref{fig:GP-by-TD-div-n-raised-third.pdf} shows the results of the macroscopic \gru{} parameters $\gamma_a = \gamma_m(\theta_a)$
from the \this{}, \toher{}, and experiment.
Here we see that the \this{} gives a consistent trend as the experiment where
diamond, zincblende, and wurtzite compounds have $\gamma_a$ mostly ranges between $0.5$ and $1.0$. For the rocksalt compounds, $\gamma_a$ 
is larger than $1.5$.
It is noticed that the \toher{} consistently overestimates the \gru{} parameters,
especially for the diamond and zincblende compounds, and this leads to low $\kappa_l$ that is seen in 
\Fig~\ref{fig:LTC-by-TD-div-n-raised-third.pdf}. 
For the diamond and zincblende compounds,
the \this{} somewhat overestimates $\kappa_l$ 
which is attributed the fact that $\theta_a$ are generally overestimated as shown
in \Fig~\ref{fig:T-acous-by-TD-div-n-raised-third.pdf}.
In contrast to these compounds, the rocksalt compounds show better agreement
between the \this{} and experiment, where  $\kappa_l$ are generally smaller than the diamond and zincblende compounds
due to larger phonon anharmonicities as 
reflected in larger \gru{} parameters for the rocksalt compounds (see \Fig~\ref{fig:GP-by-TD-div-n-raised-third.pdf}).

\begin{figure}
\centering\includegraphics[width=9.2cm,clip]{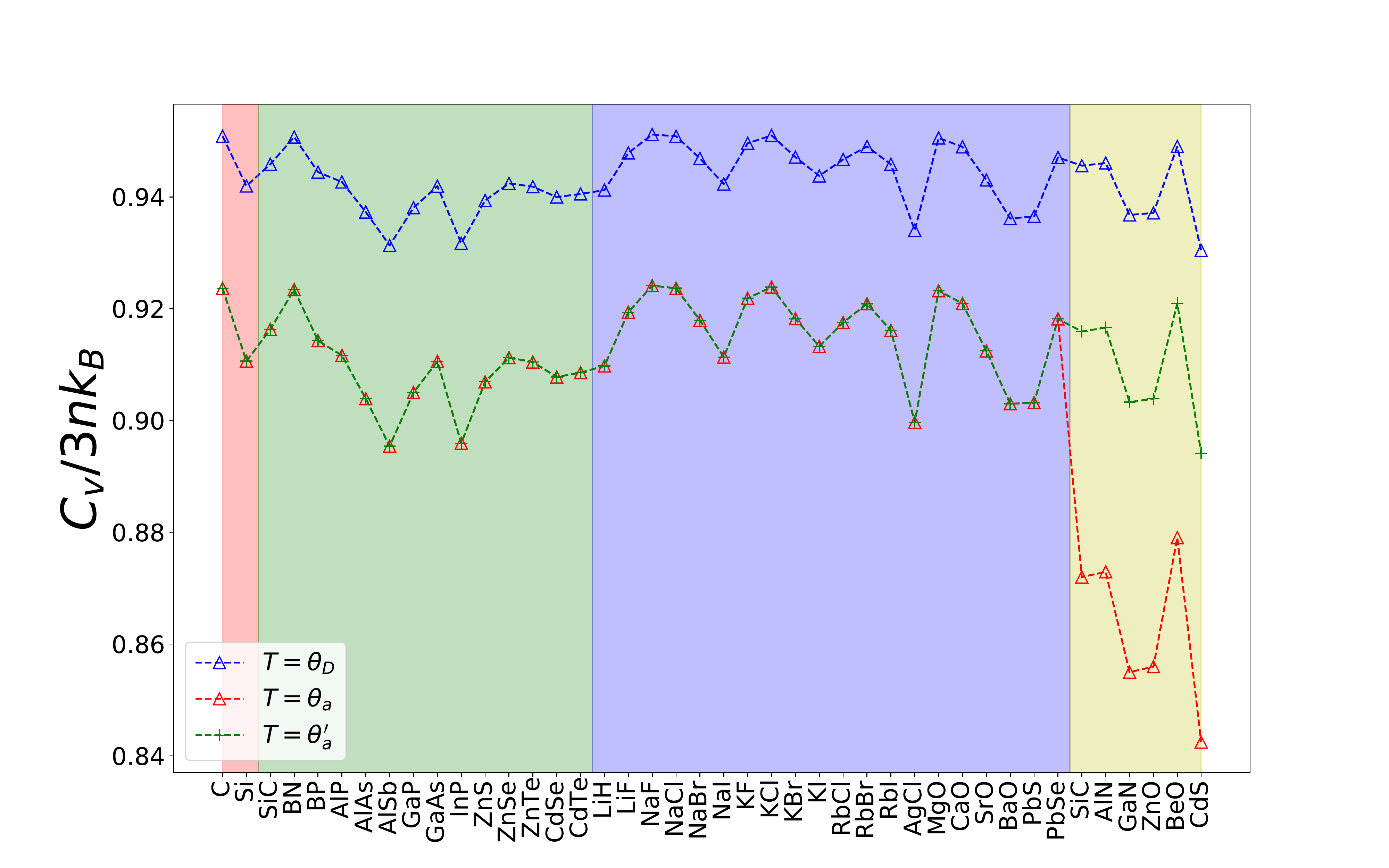}
\caption{The normalized constant-volume heat capacities for 
compounds at $T= \theta_D$, $\theta_a$ (see \Eq~\ref{eq:theta_a}), and $\theta^\prime_a$ (see \Eq~\ref{eq:modified_theta_a}).
Each compound is grouped according to the class it belongs. From the left to right, the sequence is diamond, zincblende, rocksalt, and wurtzite.
}
\label{fig:Cv-at-TD-and-Ta.pdf}
\end{figure} 

In Section~\ref{sec:method} we have discussed the possible improper estimations of $\theta_a$
that is based on \Eq~\ref{eq:theta_a}, especially when the number of atoms in the primitive cell $n$ is large.
In our study all diamond, zincblende, and rocksalt 
compounds have $n=2$, while all six wurtzite compounds SiC, AlN, GaN, ZnO, BeO, and CdS have 
$n=4$.
\Fig~\ref{fig:Cv-at-TD-and-Ta.pdf} shows the normalized heat capacity $C' = C_v/3nk_B$ evaluated 
at $\theta_D$ fluctuates
around a mean of $\sim 0.94$.  However, it is seen that $C'$ evaluated at $\theta_a$ fluctuates around
$\sim 0.91 $ and exhibits the same variation as that of $C'$ at $\theta_D$ except for the wurtzite
compounds where there is a sudden drop to lower values of $\sim 0.86$. However, $C'$ restores its expected variation
when it is evaluated using \Eq~\ref{eq:modified_theta_a}. This shows that $\theta_a$ may be more appropriately 
estimated by the same consistent fractional change
of the $\theta_D$ as suggested by \Eq~\ref{eq:modified_theta_a}. We adopt a simple modification that results in \Eq~\ref{eq:modified_theta_a}
because the original Slack formulae 
made very good predictions\cite{Morelli06} of $\kappa_l$ for compounds with $n=2$.

\begin{figure}
\centering\includegraphics[width=9.2cm,clip]{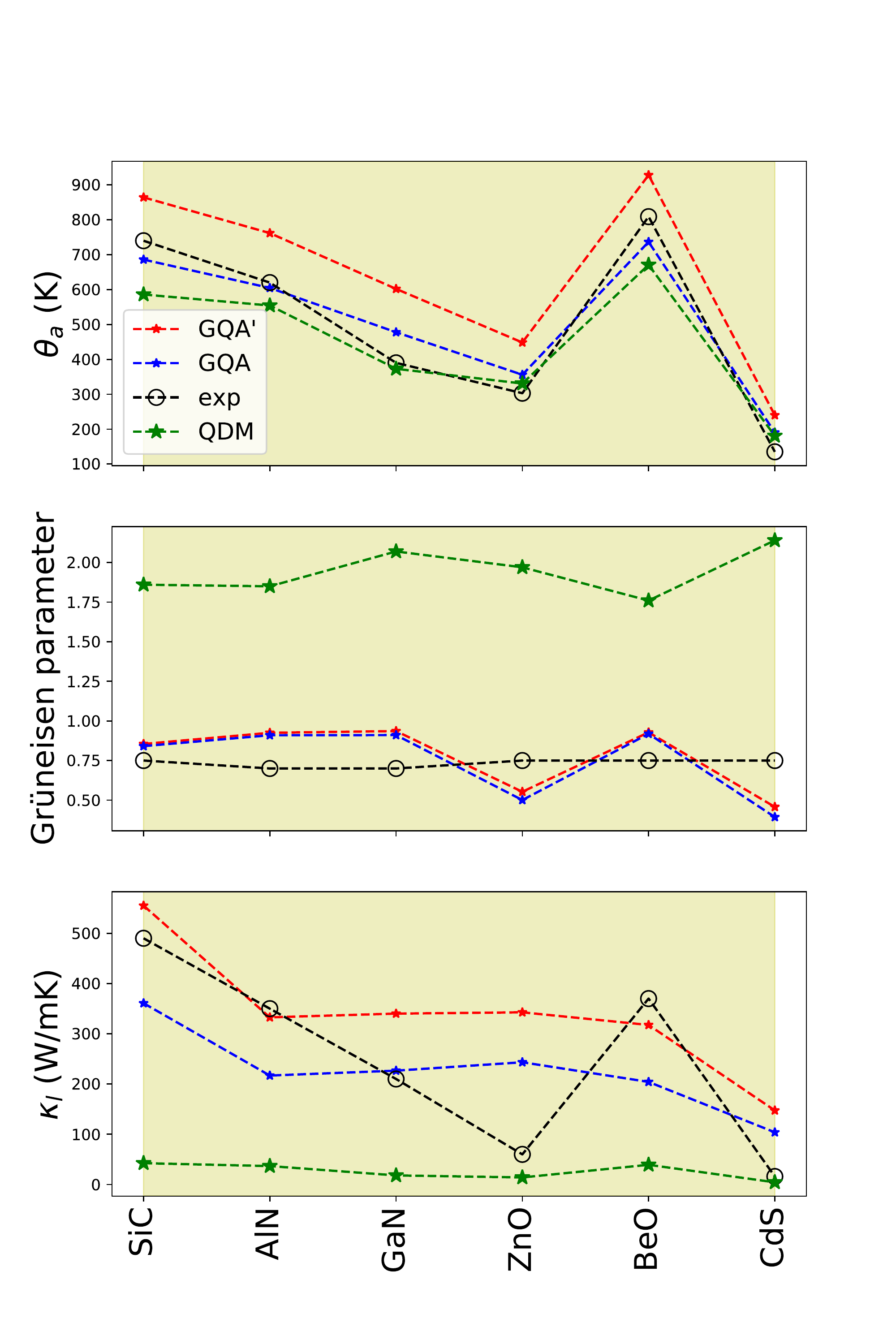}
\caption{Values of $\theta_a$, $\theta'_a$ (top panel), \gru{} parameters $\gamma_a$, $\gamma'_a$ (middle panel), $\kappa_l$, and $\kappa'_l$ (bottom panel) at $300$~K for the wurtzite compounds.
The results labelled by \this{} (\this{}') are obtained with the original (modified) Slack formulae.
}
\label{fig:thetaA-GP-kappa-for-WZ-size-consistent.pdf}
\end{figure}

For the wurtzite compounds, \Fig~\ref{fig:thetaA-GP-kappa-for-WZ-size-consistent.pdf} shows the results of the acoustic Debye temperatures $\theta_a$, $\theta'_a$
(top panel), the macroscopic \gru{} parameters $\gamma_a = \gamma_m(\theta_a)$ and $\gamma_a' = \gamma_m(\theta'_a)$ (middle panel), the lattice thermal conductivities
$\kappa_l$ and $\kappa'_a$
obtained from the original and modified 
Slack formulae (bottom panel) at $300$~K. For each of these compounds, it is not surprising that
$\theta'_a$ calculated based on the modified Slack formulae (\Eq~\ref{eq:modified_theta_a})  is predicted to be larger than that based on the
original Slack formulae (\Eq~\ref{eq:theta_a}). 
However, these values of $\theta_a$ and $\theta_a'$ make almost the same prediction
of the macroscopic \gru{} parameters. Finally, we find that $\kappa_l$ obtained with the modified
equation has a closer agreement with the experimental values for SiC, AlN, and BeO. 
These results show that the prediction of  $\kappa_l$ is rather sensitive on the way in which $\theta_a$ is 
estimated,  which is already noticeable even with a relatively small value of $n=4$. 
Future studies may include conducting comparative studies with and without the modified Slack formulae in studying compounds with large $n$ such as the      
special quasirandom structures\cite{Gan10v49} or half-Heusler crystals\cite{Shiomi11v84}.

\section{Summary}
\label{sec:summary}
In this paper we have calculated the lattice thermal 
conductivity $\kappa_l$ that is based on the original Slack formulae\cite{Morelli06} where
various equation inputs such as the Debye temperature $\theta_D$, the acoustic Debye temperature $\theta_a$, 
the \gru{} parameters are determined from the density-functional theory (DFT) phonon calculations. 
This approach, called the \gru{}-quasiharmonic approach (\this{}), is able to give $\theta_a$, \gru{} parameters, and $\kappa_l$ that follow the experimental trends well,
despite the empirical nature of the Slack formulae.
We have identified a possible size consistency problem of the original Slack formulae that involves the number of atoms in 
the primitive cell $n$. We then proposed a simple way to rectify it. 
The modified Slack formulae preserved the high efficacy of $\kappa_l$ prediction
when $n=2$, while it was shown that the effect of size consistency is noticeable even for $n$ as small as $4$ when
we considered the wurtzite structures. It is expected the \this{} with the modified Slack formulae could be used as an effective 
and practical predictor of $\kappa_l$ for a large range of materials.

\section{Acknowledgment}

We acknowledge fruitful discussions with D. Ding.
We thank the National Supercomputing Center, Singapore (NSCC) and A*STAR Computational Resource Center, 
Singapore (ACRC) for computing resources. This work is
supported by RIE2020 Advanced Manufacturing and Engineering (AME) Programmatic Grant No A1898b0043.

\bibliography{ltc}

\begin{thebibliography}{41}%
\makeatletter
\providecommand \@ifxundefined [1]{%
 \@ifx{#1\undefined}
}%
\providecommand \@ifnum [1]{%
 \ifnum #1\expandafter \@firstoftwo
 \else \expandafter \@secondoftwo
 \fi
}%
\providecommand \@ifx [1]{%
 \ifx #1\expandafter \@firstoftwo
 \else \expandafter \@secondoftwo
 \fi
}%
\providecommand \natexlab [1]{#1}%
\providecommand \enquote  [1]{``#1''}%
\providecommand \bibnamefont  [1]{#1}%
\providecommand \bibfnamefont [1]{#1}%
\providecommand \citenamefont [1]{#1}%
\providecommand \href@noop [0]{\@secondoftwo}%
\providecommand \href [0]{\begingroup \@sanitize@url \@href}%
\providecommand \@href[1]{\@@startlink{#1}\@@href}%
\providecommand \@@href[1]{\endgroup#1\@@endlink}%
\providecommand \@sanitize@url [0]{\catcode `\\12\catcode `\$12\catcode
  `\&12\catcode `\#12\catcode `\^12\catcode `\_12\catcode `\%12\relax}%
\providecommand \@@startlink[1]{}%
\providecommand \@@endlink[0]{}%
\providecommand \url  [0]{\begingroup\@sanitize@url \@url }%
\providecommand \@url [1]{\endgroup\@href {#1}{\urlprefix }}%
\providecommand \urlprefix  [0]{URL }%
\providecommand \Eprint [0]{\href }%
\providecommand \doibase [0]{https://doi.org/}%
\providecommand \selectlanguage [0]{\@gobble}%
\providecommand \bibinfo  [0]{\@secondoftwo}%
\providecommand \bibfield  [0]{\@secondoftwo}%
\providecommand \translation [1]{[#1]}%
\providecommand \BibitemOpen [0]{}%
\providecommand \bibitemStop [0]{}%
\providecommand \bibitemNoStop [0]{.\EOS\space}%
\providecommand \EOS [0]{\spacefactor3000\relax}%
\providecommand \BibitemShut  [1]{\csname bibitem#1\endcsname}%
\let\auto@bib@innerbib\@empty
\bibitem [{\citenamefont {Yan}\ and\ \citenamefont
  {Kanatzidis}(2022)}]{Yan22v21}%
  \BibitemOpen
  \bibfield  {author} {\bibinfo {author} {\bibfnamefont {Q.}~\bibnamefont
  {Yan}}\ and\ \bibinfo {author} {\bibfnamefont {M.~G.}\ \bibnamefont
  {Kanatzidis}},\ }\bibfield  {title} {\bibinfo {title} {High-performance
  thermoelectrics and challenges for practical devices},\ }\href@noop {}
  {\bibfield  {journal} {\bibinfo  {journal} {Nature Mater.}\ }\textbf
  {\bibinfo {volume} {21}},\ \bibinfo {pages} {503} (\bibinfo {year}
  {2022})}\BibitemShut {NoStop}%
\bibitem [{\citenamefont {Kundu}\ \emph {et~al.}(2021)\citenamefont {Kundu},
  \citenamefont {Yang}, \citenamefont {Ma}, \citenamefont {Feng}, \citenamefont
  {Carrete}, \citenamefont {Ruan}, \citenamefont {Madsen},\ and\ \citenamefont
  {Li}}]{Kundu21v126}%
  \BibitemOpen
  \bibfield  {author} {\bibinfo {author} {\bibfnamefont {A.}~\bibnamefont
  {Kundu}}, \bibinfo {author} {\bibfnamefont {X.}~\bibnamefont {Yang}},
  \bibinfo {author} {\bibfnamefont {J.}~\bibnamefont {Ma}}, \bibinfo {author}
  {\bibfnamefont {T.}~\bibnamefont {Feng}}, \bibinfo {author} {\bibfnamefont
  {J.}~\bibnamefont {Carrete}}, \bibinfo {author} {\bibfnamefont
  {X.}~\bibnamefont {Ruan}}, \bibinfo {author} {\bibfnamefont {G.~K.~H.}\
  \bibnamefont {Madsen}},\ and\ \bibinfo {author} {\bibfnamefont
  {W.}~\bibnamefont {Li}},\ }\bibfield  {title} {\bibinfo {title} {Ultrahigh
  thermal conductivity of $\theta$-phase tantalum nitride},\ }\href@noop {}
  {\bibfield  {journal} {\bibinfo  {journal} {Phys. Rev. Lett.}\ }\textbf
  {\bibinfo {volume} {126}},\ \bibinfo {pages} {115901} (\bibinfo {year}
  {2021})}\BibitemShut {NoStop}%
\bibitem [{\citenamefont {Chen}\ \emph {et~al.}(2019)\citenamefont {Chen},
  \citenamefont {Tran}, \citenamefont {Batra}, \citenamefont {Kim},\ and\
  \citenamefont {Ramprasad}}]{Chen19v170}%
  \BibitemOpen
  \bibfield  {author} {\bibinfo {author} {\bibfnamefont {L.}~\bibnamefont
  {Chen}}, \bibinfo {author} {\bibfnamefont {H.}~\bibnamefont {Tran}}, \bibinfo
  {author} {\bibfnamefont {R.}~\bibnamefont {Batra}}, \bibinfo {author}
  {\bibfnamefont {C.}~\bibnamefont {Kim}},\ and\ \bibinfo {author}
  {\bibfnamefont {R.}~\bibnamefont {Ramprasad}},\ }\bibfield  {title} {\bibinfo
  {title} {Machine learning models for the lattice thermal conductivity
  prediction of inorganic materials},\ }\href@noop {} {\bibfield  {journal}
  {\bibinfo  {journal} {Comput. Mater. Sci.}\ }\textbf {\bibinfo {volume}
  {170}},\ \bibinfo {pages} {109155} (\bibinfo {year} {2019})}\BibitemShut
  {NoStop}%
\bibitem [{\citenamefont {Miller}\ \emph {et~al.}(2017)\citenamefont {Miller},
  \citenamefont {Gorai}, \citenamefont {Ortiz}, \citenamefont {Goyal},
  \citenamefont {Gao}, \citenamefont {Barnett}, \citenamefont {Mason},
  \citenamefont {Lv}, \citenamefont {Stevanovic},\ and\ \citenamefont
  {Toberer}}]{Miller17v29}%
  \BibitemOpen
  \bibfield  {author} {\bibinfo {author} {\bibfnamefont {S.~A.}\ \bibnamefont
  {Miller}}, \bibinfo {author} {\bibfnamefont {P.}~\bibnamefont {Gorai}},
  \bibinfo {author} {\bibfnamefont {B.~R.}\ \bibnamefont {Ortiz}}, \bibinfo
  {author} {\bibfnamefont {A.}~\bibnamefont {Goyal}}, \bibinfo {author}
  {\bibfnamefont {D.}~\bibnamefont {Gao}}, \bibinfo {author} {\bibfnamefont
  {S.~A.}\ \bibnamefont {Barnett}}, \bibinfo {author} {\bibfnamefont {T.~O.}\
  \bibnamefont {Mason}}, \bibinfo {author} {\bibfnamefont {G.~J. S.~Q.}\
  \bibnamefont {Lv}}, \bibinfo {author} {\bibfnamefont {V.}~\bibnamefont
  {Stevanovic}},\ and\ \bibinfo {author} {\bibfnamefont {E.~S.}\ \bibnamefont
  {Toberer}},\ }\bibfield  {title} {\bibinfo {title} {Capturing anharmonicity
  in a lattice thermal conductivity model for high-throughput predictions},\
  }\href@noop {} {\bibfield  {journal} {\bibinfo  {journal} {Chem. Mater.}\
  }\textbf {\bibinfo {volume} {29}},\ \bibinfo {pages} {2494} (\bibinfo {year}
  {2017})}\BibitemShut {NoStop}%
\bibitem [{\citenamefont {Yu}\ \emph {et~al.}(2016)\citenamefont {Yu},
  \citenamefont {Dai},\ and\ \citenamefont {Chen}}]{Yu16v6}%
  \BibitemOpen
  \bibfield  {author} {\bibinfo {author} {\bibfnamefont {H.}~\bibnamefont
  {Yu}}, \bibinfo {author} {\bibfnamefont {S.}~\bibnamefont {Dai}},\ and\
  \bibinfo {author} {\bibfnamefont {Y.}~\bibnamefont {Chen}},\ }\bibfield
  {title} {\bibinfo {title} {Enhanced power factor via the control of
  structural phase transition in {S}n{S}e},\ }\href@noop {} {\bibfield
  {journal} {\bibinfo  {journal} {Sci. Rep.}\ }\textbf {\bibinfo {volume}
  {6}},\ \bibinfo {pages} {1} (\bibinfo {year} {2016})}\BibitemShut {NoStop}%
\bibitem [{\citenamefont {Waldrop}(2016)}]{Waldrop16v530}%
  \BibitemOpen
  \bibfield  {author} {\bibinfo {author} {\bibfnamefont {M.~M.}\ \bibnamefont
  {Waldrop}},\ }\bibfield  {title} {\bibinfo {title} {More than {M}oore},\
  }\href@noop {} {\bibfield  {journal} {\bibinfo  {journal} {Nature}\ }\textbf
  {\bibinfo {volume} {530}},\ \bibinfo {pages} {145} (\bibinfo {year}
  {2016})}\BibitemShut {NoStop}%
\bibitem [{\citenamefont {Nath}\ \emph {et~al.}(2017)\citenamefont {Nath},
  \citenamefont {Plata}, \citenamefont {Usanmaz}, \citenamefont {Toher},
  \citenamefont {Fornari}, \citenamefont {Nardelli},\ and\ \citenamefont
  {Curtarolo}}]{Nath17v129}%
  \BibitemOpen
  \bibfield  {author} {\bibinfo {author} {\bibfnamefont {P.}~\bibnamefont
  {Nath}}, \bibinfo {author} {\bibfnamefont {J.~J.}\ \bibnamefont {Plata}},
  \bibinfo {author} {\bibfnamefont {D.}~\bibnamefont {Usanmaz}}, \bibinfo
  {author} {\bibfnamefont {C.}~\bibnamefont {Toher}}, \bibinfo {author}
  {\bibfnamefont {M.}~\bibnamefont {Fornari}}, \bibinfo {author} {\bibfnamefont
  {M.~B.}\ \bibnamefont {Nardelli}},\ and\ \bibinfo {author} {\bibfnamefont
  {S.}~\bibnamefont {Curtarolo}},\ }\bibfield  {title} {\bibinfo {title} {High
  throughput combinatorial method for fast and robust prediction of lattice
  thermal conductivity},\ }\href@noop {} {\bibfield  {journal} {\bibinfo
  {journal} {Scr. Mater.}\ }\textbf {\bibinfo {volume} {129}},\ \bibinfo
  {pages} {88} (\bibinfo {year} {2017})}\BibitemShut {NoStop}%
\bibitem [{\citenamefont {Tadano}\ and\ \citenamefont
  {Tsuneyuki}(2018)}]{Tadano18v120}%
  \BibitemOpen
  \bibfield  {author} {\bibinfo {author} {\bibfnamefont {T.}~\bibnamefont
  {Tadano}}\ and\ \bibinfo {author} {\bibfnamefont {S.}~\bibnamefont
  {Tsuneyuki}},\ }\bibfield  {title} {\bibinfo {title} {Quartic anharnonicity
  of rattlers and its effect on lattice thermal conductivity of clathrates from
  first principles},\ }\href@noop {} {\bibfield  {journal} {\bibinfo  {journal}
  {Phys. Rev. Lett.}\ }\textbf {\bibinfo {volume} {120}},\ \bibinfo {pages}
  {105901} (\bibinfo {year} {2018})}\BibitemShut {NoStop}%
\bibitem [{\citenamefont {Seko}\ \emph {et~al.}(2015)\citenamefont {Seko},
  \citenamefont {Togo}, \citenamefont {Hayashi}, \citenamefont {Tsuda},
  \citenamefont {Chaput},\ and\ \citenamefont {Tanaka}}]{Seko15v115}%
  \BibitemOpen
  \bibfield  {author} {\bibinfo {author} {\bibfnamefont {A.}~\bibnamefont
  {Seko}}, \bibinfo {author} {\bibfnamefont {A.}~\bibnamefont {Togo}}, \bibinfo
  {author} {\bibfnamefont {H.}~\bibnamefont {Hayashi}}, \bibinfo {author}
  {\bibfnamefont {K.}~\bibnamefont {Tsuda}}, \bibinfo {author} {\bibfnamefont
  {L.}~\bibnamefont {Chaput}},\ and\ \bibinfo {author} {\bibfnamefont
  {I.}~\bibnamefont {Tanaka}},\ }\bibfield  {title} {\bibinfo {title}
  {Prediction of low thermal conductivity compounds with first-principles
  anharmonic lattice-dynamics calculations and bayesian optimization},\
  }\href@noop {} {\bibfield  {journal} {\bibinfo  {journal} {Phys. Rev. Lett.}\
  }\textbf {\bibinfo {volume} {2}},\ \bibinfo {pages} {17053} (\bibinfo {year}
  {2015})}\BibitemShut {NoStop}%
\bibitem [{\citenamefont {Lindsay}(2016)}]{Lindsay16v20}%
  \BibitemOpen
  \bibfield  {author} {\bibinfo {author} {\bibfnamefont {L.}~\bibnamefont
  {Lindsay}},\ }\bibfield  {title} {\bibinfo {title} {First principles
  {P}eierls-{B}oltzmann phonon thermal transport: A topical review},\
  }\href@noop {} {\bibfield  {journal} {\bibinfo  {journal} {Nano. Micro.
  Thermophy. Eng.}\ }\textbf {\bibinfo {volume} {20}},\ \bibinfo {pages} {67}
  (\bibinfo {year} {2016})}\BibitemShut {NoStop}%
\bibitem [{\citenamefont {Ward}\ \emph {et~al.}(2009)\citenamefont {Ward},
  \citenamefont {Broido}, \citenamefont {Stewart},\ and\ \citenamefont
  {Deinzer}}]{Ward09v80}%
  \BibitemOpen
  \bibfield  {author} {\bibinfo {author} {\bibfnamefont {A.}~\bibnamefont
  {Ward}}, \bibinfo {author} {\bibfnamefont {D.~A.}\ \bibnamefont {Broido}},
  \bibinfo {author} {\bibfnamefont {D.~A.}\ \bibnamefont {Stewart}},\ and\
  \bibinfo {author} {\bibfnamefont {G.}~\bibnamefont {Deinzer}},\ }\bibfield
  {title} {\bibinfo {title} {Ab initio theory of the lattice thermal
  conductivity in diamond},\ }\href@noop {} {\bibfield  {journal} {\bibinfo
  {journal} {Phys. Rev. B}\ }\textbf {\bibinfo {volume} {80}},\ \bibinfo
  {pages} {125203} (\bibinfo {year} {2009})}\BibitemShut {NoStop}%
\bibitem [{\citenamefont {Toher}\ \emph {et~al.}(2014)\citenamefont {Toher},
  \citenamefont {Plata}, \citenamefont {Levy}, \citenamefont {de~Jong},
  \citenamefont {Asta}, \citenamefont {BuongiornoNardelli},\ and\ \citenamefont
  {Curtarolo}}]{Toher14v90}%
  \BibitemOpen
  \bibfield  {author} {\bibinfo {author} {\bibfnamefont {C.}~\bibnamefont
  {Toher}}, \bibinfo {author} {\bibfnamefont {J.~J.}\ \bibnamefont {Plata}},
  \bibinfo {author} {\bibfnamefont {O.}~\bibnamefont {Levy}}, \bibinfo {author}
  {\bibfnamefont {M.}~\bibnamefont {de~Jong}}, \bibinfo {author} {\bibfnamefont
  {M.}~\bibnamefont {Asta}}, \bibinfo {author} {\bibfnamefont {M.}~\bibnamefont
  {BuongiornoNardelli}},\ and\ \bibinfo {author} {\bibfnamefont
  {S.}~\bibnamefont {Curtarolo}},\ }\bibfield  {title} {\bibinfo {title}
  {High-throughput computational screening of thermal conductivity, {D}ebye
  temperature, and {G}r\"uneisen parameter using a quasiharmonic {D}ebye
  model},\ }\href@noop {} {\bibfield  {journal} {\bibinfo  {journal} {Phys.
  Rev. B}\ }\textbf {\bibinfo {volume} {90}},\ \bibinfo {pages} {174107}
  (\bibinfo {year} {2014})}\BibitemShut {NoStop}%
\bibitem [{\citenamefont {Bjerg}\ \emph {et~al.}(2014)\citenamefont {Bjerg},
  \citenamefont {Iversen},\ and\ \citenamefont {Madsen}}]{Bjerg14v89}%
  \BibitemOpen
  \bibfield  {author} {\bibinfo {author} {\bibfnamefont {L.}~\bibnamefont
  {Bjerg}}, \bibinfo {author} {\bibfnamefont {B.~B.}\ \bibnamefont {Iversen}},\
  and\ \bibinfo {author} {\bibfnamefont {G.~K.~H.}\ \bibnamefont {Madsen}},\
  }\bibfield  {title} {\bibinfo {title} {Modeling the thermal conductivities of
  the zinc antimonides {Z}n{S}b and {Z}n$_4${S}b$_3$},\ }\href@noop {}
  {\bibfield  {journal} {\bibinfo  {journal} {Phys. Rev. B}\ }\textbf {\bibinfo
  {volume} {89}},\ \bibinfo {pages} {024304} (\bibinfo {year}
  {2014})}\BibitemShut {NoStop}%
\bibitem [{\citenamefont {Gorai}\ \emph {et~al.}(2017)\citenamefont {Gorai},
  \citenamefont {Stevanovi\'c},\ and\ \citenamefont {Toberer}}]{Gorai17v2}%
  \BibitemOpen
  \bibfield  {author} {\bibinfo {author} {\bibfnamefont {P.}~\bibnamefont
  {Gorai}}, \bibinfo {author} {\bibfnamefont {V.}~\bibnamefont
  {Stevanovi\'c}},\ and\ \bibinfo {author} {\bibfnamefont {E.}~\bibnamefont
  {Toberer}},\ }\bibfield  {title} {\bibinfo {title} {Computationally guided
  discovery of thermoelectric materials},\ }\href@noop {} {\bibfield  {journal}
  {\bibinfo  {journal} {Nature Rev. Mat.}\ }\textbf {\bibinfo {volume} {2}},\
  \bibinfo {pages} {17053} (\bibinfo {year} {2017})}\BibitemShut {NoStop}%
\bibitem [{\citenamefont {Morelli}\ and\ \citenamefont
  {Slack}(2006)}]{Morelli06}%
  \BibitemOpen
  \bibfield  {author} {\bibinfo {author} {\bibfnamefont {D.~T.}\ \bibnamefont
  {Morelli}}\ and\ \bibinfo {author} {\bibfnamefont {G.~A.}\ \bibnamefont
  {Slack}},\ }\bibfield  {title} {\bibinfo {title} {High lattice thermal
  conductivity solids},\ }in\ \href@noop {} {\emph {\bibinfo {booktitle} {High
  Thermal Conductivity Materials}}},\ \bibinfo {editor} {edited by\ \bibinfo
  {editor} {\bibfnamefont {S.~L.}\ \bibnamefont {Shind\'e}}\ and\ \bibinfo
  {editor} {\bibfnamefont {J.~S.}\ \bibnamefont {Goela}}}\ (\bibinfo
  {publisher} {Springer, New York},\ \bibinfo {year} {2006})\ p.~\bibinfo
  {pages} {37}\BibitemShut {NoStop}%
\bibitem [{\citenamefont {Blanco}\ \emph {et~al.}(2004)\citenamefont {Blanco},
  \citenamefont {Francisco},\ and\ \citenamefont {Lua{\~n}a}}]{Blanco04v57}%
  \BibitemOpen
  \bibfield  {author} {\bibinfo {author} {\bibfnamefont {M.~A.}\ \bibnamefont
  {Blanco}}, \bibinfo {author} {\bibfnamefont {E.}~\bibnamefont {Francisco}},\
  and\ \bibinfo {author} {\bibfnamefont {V.}~\bibnamefont {Lua{\~n}a}},\
  }\bibfield  {title} {\bibinfo {title} {Gibbs: isothermal-isobaric
  thermodynamics of solids from energy curves using a quasi-harmonic debye
  model},\ }\href@noop {} {\bibfield  {journal} {\bibinfo  {journal} {Comput.
  Phys. Commun.}\ }\textbf {\bibinfo {volume} {158}},\ \bibinfo {pages} {57}
  (\bibinfo {year} {2004})}\BibitemShut {NoStop}%
\bibitem [{\citenamefont {Garg}\ \emph {et~al.}(2011)\citenamefont {Garg},
  \citenamefont {Bonin}, \citenamefont {Kozinsky},\ and\ \citenamefont
  {Marzari}}]{Garg11v106}%
  \BibitemOpen
  \bibfield  {author} {\bibinfo {author} {\bibfnamefont {J.}~\bibnamefont
  {Garg}}, \bibinfo {author} {\bibfnamefont {N.}~\bibnamefont {Bonin}},
  \bibinfo {author} {\bibfnamefont {B.}~\bibnamefont {Kozinsky}},\ and\
  \bibinfo {author} {\bibfnamefont {N.}~\bibnamefont {Marzari}},\ }\bibfield
  {title} {\bibinfo {title} {Role of disorder and anharmonicity in the thermal
  conductivity of silicon-germanium alloys: A first-principles study},\
  }\href@noop {} {\bibfield  {journal} {\bibinfo  {journal} {Phys. Rev. Lett.}\
  }\textbf {\bibinfo {volume} {106}},\ \bibinfo {pages} {045901} (\bibinfo
  {year} {2011})}\BibitemShut {NoStop}%
\bibitem [{Mat(2022)}]{MatProj-link}%
  \BibitemOpen
  \href@noop {} {\bibinfo {title} {The {M}aterials {P}roject}},\ \bibinfo
  {howpublished} {\url{https://materialsproject.org}} (\bibinfo {year}
  {2022})\BibitemShut {NoStop}%
\bibitem [{\citenamefont {Petretto}\ \emph {et~al.}(2018)\citenamefont
  {Petretto}, \citenamefont {Dwaraknath}, \citenamefont {Miranda},
  \citenamefont {Winston}, \citenamefont {Giantomassi}, \citenamefont {van
  Setten}, \citenamefont {Gonze}, \citenamefont {Persson}, \citenamefont
  {Hautier},\ and\ \citenamefont {Rignanese}}]{Petretto18v5}%
  \BibitemOpen
  \bibfield  {author} {\bibinfo {author} {\bibfnamefont {G.}~\bibnamefont
  {Petretto}}, \bibinfo {author} {\bibfnamefont {S.}~\bibnamefont
  {Dwaraknath}}, \bibinfo {author} {\bibfnamefont {H.~P.}\ \bibnamefont
  {Miranda}}, \bibinfo {author} {\bibfnamefont {D.}~\bibnamefont {Winston}},
  \bibinfo {author} {\bibfnamefont {M.}~\bibnamefont {Giantomassi}}, \bibinfo
  {author} {\bibfnamefont {M.~J.}\ \bibnamefont {van Setten}}, \bibinfo
  {author} {\bibfnamefont {X.}~\bibnamefont {Gonze}}, \bibinfo {author}
  {\bibfnamefont {K.~A.}\ \bibnamefont {Persson}}, \bibinfo {author}
  {\bibfnamefont {G.}~\bibnamefont {Hautier}},\ and\ \bibinfo {author}
  {\bibfnamefont {G.-M.}\ \bibnamefont {Rignanese}},\ }\bibfield  {title}
  {\bibinfo {title} {Data descriptor: High-throughput density-functional
  perturbation theory phonons for inorganic materials},\ }\href@noop {}
  {\bibfield  {journal} {\bibinfo  {journal} {Sci. Data}\ }\textbf {\bibinfo
  {volume} {5}},\ \bibinfo {pages} {180065} (\bibinfo {year}
  {2018})}\BibitemShut {NoStop}%
\bibitem [{\citenamefont {Togo}(2020)}]{TogoPhononDB2020-link}%
  \BibitemOpen
  \bibfield  {author} {\bibinfo {author} {\bibfnamefont {A.}~\bibnamefont
  {Togo}},\ }\href@noop {} {\bibinfo {title} {Phonon database at {K}yoto
  {U}niversity}},\ \bibinfo {howpublished}
  {\url{http://phonondb.mtl.kyoto-u.ac.jp}} (\bibinfo {year}
  {2020})\BibitemShut {NoStop}%
\bibitem [{\citenamefont {Schelling}\ and\ \citenamefont
  {Keblinski}(2003)}]{Schelling03v68}%
  \BibitemOpen
  \bibfield  {author} {\bibinfo {author} {\bibfnamefont {P.~K.}\ \bibnamefont
  {Schelling}}\ and\ \bibinfo {author} {\bibfnamefont {P.}~\bibnamefont
  {Keblinski}},\ }\bibfield  {title} {\bibinfo {title} {Thermal expansion of
  carbon structures},\ }\href@noop {} {\bibfield  {journal} {\bibinfo
  {journal} {Phys. Rev. B}\ }\textbf {\bibinfo {volume} {68}},\ \bibinfo
  {pages} {035425} (\bibinfo {year} {2003})}\BibitemShut {NoStop}%
\bibitem [{\citenamefont {Ding}\ and\ \citenamefont {Xiao}(2015)}]{Ding15v5}%
  \BibitemOpen
  \bibfield  {author} {\bibinfo {author} {\bibfnamefont {Y.}~\bibnamefont
  {Ding}}\ and\ \bibinfo {author} {\bibfnamefont {B.}~\bibnamefont {Xiao}},\
  }\bibfield  {title} {\bibinfo {title} {Thermal expansion tensors,
  {G}r{\"u}neisen parameters and phonon velocities of bulk {MT2} ({M=W} and
  {M}o; {T}={S} and {S}e) from first principles calculations},\ }\href@noop {}
  {\bibfield  {journal} {\bibinfo  {journal} {RSC Adv.}\ }\textbf {\bibinfo
  {volume} {5}},\ \bibinfo {pages} {18391} (\bibinfo {year}
  {2015})}\BibitemShut {NoStop}%
\bibitem [{\citenamefont {Gan}\ \emph {et~al.}(2015)\citenamefont {Gan},
  \citenamefont {Soh},\ and\ \citenamefont {Liu}}]{Gan15v92}%
  \BibitemOpen
  \bibfield  {author} {\bibinfo {author} {\bibfnamefont {C.~K.}\ \bibnamefont
  {Gan}}, \bibinfo {author} {\bibfnamefont {J.~R.}\ \bibnamefont {Soh}},\ and\
  \bibinfo {author} {\bibfnamefont {Y.}~\bibnamefont {Liu}},\ }\bibfield
  {title} {\bibinfo {title} {Large anharmonic effect and thermal expansion
  anisotropy of metal chalcogenides: The case of antimony sulfide},\
  }\href@noop {} {\bibfield  {journal} {\bibinfo  {journal} {Phys. Rev. B}\
  }\textbf {\bibinfo {volume} {92}},\ \bibinfo {pages} {235202} (\bibinfo
  {year} {2015})}\BibitemShut {NoStop}%
\bibitem [{\citenamefont {Liu}\ \emph {et~al.}(2017)\citenamefont {Liu},
  \citenamefont {Liu}, \citenamefont {Zhou},\ and\ \citenamefont
  {Wan}}]{Liu17v121}%
  \BibitemOpen
  \bibfield  {author} {\bibinfo {author} {\bibfnamefont {G.}~\bibnamefont
  {Liu}}, \bibinfo {author} {\bibfnamefont {H.~M.}\ \bibnamefont {Liu}},
  \bibinfo {author} {\bibfnamefont {J.}~\bibnamefont {Zhou}},\ and\ \bibinfo
  {author} {\bibfnamefont {X.~G.}\ \bibnamefont {Wan}},\ }\bibfield  {title}
  {\bibinfo {title} {Temperature effect on lattice and electronic structures of
  {W}{T}e$_2$ from first-principles study},\ }\href@noop {} {\bibfield
  {journal} {\bibinfo  {journal} {J. Appl. Phys.}\ }\textbf {\bibinfo {volume}
  {121}},\ \bibinfo {pages} {045104} (\bibinfo {year} {2017})}\BibitemShut
  {NoStop}%
\bibitem [{\citenamefont {Gan}\ and\ \citenamefont {Lee}(2018)}]{Gan18v151}%
  \BibitemOpen
  \bibfield  {author} {\bibinfo {author} {\bibfnamefont {C.~K.}\ \bibnamefont
  {Gan}}\ and\ \bibinfo {author} {\bibfnamefont {C.~H.}\ \bibnamefont {Lee}},\
  }\bibfield  {title} {\bibinfo {title} {Anharmonic phonon effects on linear
  thermal expansion of trigonal bismuth selenide and antimony telluride
  crystals},\ }\href@noop {} {\bibfield  {journal} {\bibinfo  {journal}
  {Comput. Mater. Sci.}\ }\textbf {\bibinfo {volume} {151}},\ \bibinfo {pages}
  {49} (\bibinfo {year} {2018})}\BibitemShut {NoStop}%
\bibitem [{\citenamefont {Liu}\ and\ \citenamefont {Allen}(2018)}]{Liu18v154}%
  \BibitemOpen
  \bibfield  {author} {\bibinfo {author} {\bibfnamefont {J.}~\bibnamefont
  {Liu}}\ and\ \bibinfo {author} {\bibfnamefont {P.~B.}\ \bibnamefont
  {Allen}},\ }\bibfield  {title} {\bibinfo {title} {Internal and external
  expansions of wurtzite {Z}n{O} from first principles},\ }\href@noop {}
  {\bibfield  {journal} {\bibinfo  {journal} {Comput. Mater. Sci.}\ }\textbf
  {\bibinfo {volume} {154}},\ \bibinfo {pages} {251} (\bibinfo {year}
  {2018})}\BibitemShut {NoStop}%
\bibitem [{\citenamefont {Gan}\ and\ \citenamefont {Chua}(2019)}]{Gan19v31}%
  \BibitemOpen
  \bibfield  {author} {\bibinfo {author} {\bibfnamefont {C.~K.}\ \bibnamefont
  {Gan}}\ and\ \bibinfo {author} {\bibfnamefont {K.~T.~E.}\ \bibnamefont
  {Chua}},\ }\bibfield  {title} {\bibinfo {title} {Large thermal anisotropy in
  monoclinic niobium trisulfide: A thermal expansion tensor study},\
  }\href@noop {} {\bibfield  {journal} {\bibinfo  {journal} {J. Phys.: Condens.
  Matter}\ }\textbf {\bibinfo {volume} {31}},\ \bibinfo {pages} {265401}
  (\bibinfo {year} {2019})}\BibitemShut {NoStop}%
\bibitem [{\citenamefont {Liu}\ \emph {et~al.}(2019)\citenamefont {Liu},
  \citenamefont {Gao},\ and\ \citenamefont {Ren}}]{Liu19v99}%
  \BibitemOpen
  \bibfield  {author} {\bibinfo {author} {\bibfnamefont {G.}~\bibnamefont
  {Liu}}, \bibinfo {author} {\bibfnamefont {Z.}~\bibnamefont {Gao}},\ and\
  \bibinfo {author} {\bibfnamefont {J.}~\bibnamefont {Ren}},\ }\bibfield
  {title} {\bibinfo {title} {Anisotropic thermal expansion and thermodynamic
  properties of monolayer beta-{T}e},\ }\href@noop {} {\bibfield  {journal}
  {\bibinfo  {journal} {Phys. Rev. B}\ }\textbf {\bibinfo {volume} {99}},\
  \bibinfo {pages} {195436} (\bibinfo {year} {2019})}\BibitemShut {NoStop}%
\bibitem [{\citenamefont {Gan}\ \emph {et~al.}(2022)\citenamefont {Gan},
  \citenamefont {Al-Sharif}, \citenamefont {Al-Shorman},\ and\ \citenamefont
  {Qteish}}]{Gan22v12}%
  \BibitemOpen
  \bibfield  {author} {\bibinfo {author} {\bibfnamefont {C.~K.}\ \bibnamefont
  {Gan}}, \bibinfo {author} {\bibfnamefont {A.~I.}\ \bibnamefont {Al-Sharif}},
  \bibinfo {author} {\bibfnamefont {A.}~\bibnamefont {Al-Shorman}},\ and\
  \bibinfo {author} {\bibfnamefont {A.}~\bibnamefont {Qteish}},\ }\bibfield
  {title} {\bibinfo {title} {A first-principles investigation of the linear
  thermal expansion coefficients of {B}e{F}$_2$: Giant thermal expansion},\
  }\href@noop {} {\bibfield  {journal} {\bibinfo  {journal} {RSC Adv.}\
  }\textbf {\bibinfo {volume} {12}},\ \bibinfo {pages} {26588} (\bibinfo {year}
  {2022})}\BibitemShut {NoStop}%
\bibitem [{\citenamefont {Leibfried}\ and\ \citenamefont
  {Schl\"omann}(1954)}]{Leibfried54v71}%
  \BibitemOpen
  \bibfield  {author} {\bibinfo {author} {\bibfnamefont {G.}~\bibnamefont
  {Leibfried}}\ and\ \bibinfo {author} {\bibfnamefont {E.}~\bibnamefont
  {Schl\"omann}},\ }\href@noop {} {\bibfield  {journal} {\bibinfo  {journal}
  {Nachr. Akad. Wiss. G\"ottinger II}\ }\textbf {\bibinfo {volume} {a(4)}},\
  \bibinfo {pages} {71} (\bibinfo {year} {1954})}\BibitemShut {NoStop}%
\bibitem [{\citenamefont {Julian}(1965)}]{Julian65v137}%
  \BibitemOpen
  \bibfield  {author} {\bibinfo {author} {\bibfnamefont {C.~L.}\ \bibnamefont
  {Julian}},\ }\bibfield  {title} {\bibinfo {title} {Theory of heat conduction
  in rare-gas crystals},\ }\href@noop {} {\bibfield  {journal} {\bibinfo
  {journal} {Phys. Rev.}\ }\textbf {\bibinfo {volume} {137}},\ \bibinfo {pages}
  {A128} (\bibinfo {year} {1965})}\BibitemShut {NoStop}%
\bibitem [{\citenamefont {Slack}(1979)}]{Slack79v34}%
  \BibitemOpen
  \bibfield  {author} {\bibinfo {author} {\bibfnamefont {G.~A.}\ \bibnamefont
  {Slack}},\ }in\ \href@noop {} {\emph {\bibinfo {booktitle} {Solid State
  Physics}}},\ Vol.~\bibinfo {volume} {34},\ \bibinfo {editor} {edited by\
  \bibinfo {editor} {\bibfnamefont {H.}~\bibnamefont {Ehrenreich}}, \bibinfo
  {editor} {\bibfnamefont {F.}~\bibnamefont {Seitz}},\ and\ \bibinfo {editor}
  {\bibfnamefont {D.}~\bibnamefont {Turnbull}}}\ (\bibinfo  {publisher}
  {Academic, New York},\ \bibinfo {year} {1979})\ p.~\bibinfo {pages}
  {1}\BibitemShut {NoStop}%
\bibitem [{\citenamefont {McQuarrie}(2000)}]{McQuarrie2000-book}%
  \BibitemOpen
  \bibfield  {author} {\bibinfo {author} {\bibfnamefont {D.~A.}\ \bibnamefont
  {McQuarrie}},\ }\href@noop {} {\emph {\bibinfo {title} {Statistical
  Mechanics}}}\ (\bibinfo  {publisher} {University Science Books},\ \bibinfo
  {address} {Sausalito},\ \bibinfo {year} {2000})\BibitemShut {NoStop}%
\bibitem [{\citenamefont {Ashcroft}\ and\ \citenamefont
  {Mermin}(1976)}]{AshcroftMermin1976-book}%
  \BibitemOpen
  \bibfield  {author} {\bibinfo {author} {\bibfnamefont {N.~W.}\ \bibnamefont
  {Ashcroft}}\ and\ \bibinfo {author} {\bibfnamefont {N.~D.}\ \bibnamefont
  {Mermin}},\ }\href@noop {} {\emph {\bibinfo {title} {Solid State Physics}}}\
  (\bibinfo  {publisher} {Saunders College Publishing},\ \bibinfo {address}
  {New York},\ \bibinfo {year} {1976})\BibitemShut {NoStop}%
\bibitem [{\citenamefont {Kresse}\ and\ \citenamefont
  {Furthm{\"u}ller}(1996)}]{Kresse96v6}%
  \BibitemOpen
  \bibfield  {author} {\bibinfo {author} {\bibfnamefont {G.}~\bibnamefont
  {Kresse}}\ and\ \bibinfo {author} {\bibfnamefont {J.}~\bibnamefont
  {Furthm{\"u}ller}},\ }\bibfield  {title} {\bibinfo {title} {Efficiency of
  ab-initio total energy calculations for metals and semiconductors using a
  plane-wave basis set},\ }\href@noop {} {\bibfield  {journal} {\bibinfo
  {journal} {Comput. Mater. Sci.}\ }\textbf {\bibinfo {volume} {6}},\ \bibinfo
  {pages} {15} (\bibinfo {year} {1996})}\BibitemShut {NoStop}%
\bibitem [{\citenamefont {Perdew}\ \emph {et~al.}(2008)\citenamefont {Perdew},
  \citenamefont {Ruzsinszky}, \citenamefont {Csonka}, \citenamefont {Vydrov},
  \citenamefont {Scuseria}, \citenamefont {Constantin}, \citenamefont {Zhou},\
  and\ \citenamefont {Burke}}]{Perdew08v100}%
  \BibitemOpen
  \bibfield  {author} {\bibinfo {author} {\bibfnamefont {J.~P.}\ \bibnamefont
  {Perdew}}, \bibinfo {author} {\bibfnamefont {A.}~\bibnamefont {Ruzsinszky}},
  \bibinfo {author} {\bibfnamefont {G.~I.}\ \bibnamefont {Csonka}}, \bibinfo
  {author} {\bibfnamefont {O.~A.}\ \bibnamefont {Vydrov}}, \bibinfo {author}
  {\bibfnamefont {G.~E.}\ \bibnamefont {Scuseria}}, \bibinfo {author}
  {\bibfnamefont {L.~A.}\ \bibnamefont {Constantin}}, \bibinfo {author}
  {\bibfnamefont {X.}~\bibnamefont {Zhou}},\ and\ \bibinfo {author}
  {\bibfnamefont {K.}~\bibnamefont {Burke}},\ }\bibfield  {title} {\bibinfo
  {title} {Restoring the density-gradient expansion for exchange in solids and
  surfaces},\ }\href@noop {} {\bibfield  {journal} {\bibinfo  {journal} {Phys.
  Rev. Lett.}\ }\textbf {\bibinfo {volume} {100}},\ \bibinfo {pages} {136406}
  (\bibinfo {year} {2008})}\BibitemShut {NoStop}%
\bibitem [{\citenamefont {Kresse}\ \emph {et~al.}(1995)\citenamefont {Kresse},
  \citenamefont {Furthm\"uller},\ and\ \citenamefont {Hafner}}]{Kresse95v32}%
  \BibitemOpen
  \bibfield  {author} {\bibinfo {author} {\bibfnamefont {G.}~\bibnamefont
  {Kresse}}, \bibinfo {author} {\bibfnamefont {J.}~\bibnamefont
  {Furthm\"uller}},\ and\ \bibinfo {author} {\bibfnamefont {J.}~\bibnamefont
  {Hafner}},\ }\bibfield  {title} {\bibinfo {title} {Ab initio force constant
  approach to phonon dispersion relations of diamond and graphite},\
  }\href@noop {} {\bibfield  {journal} {\bibinfo  {journal} {Europhys. Lett.}\
  }\textbf {\bibinfo {volume} {32}},\ \bibinfo {pages} {729} (\bibinfo {year}
  {1995})}\BibitemShut {NoStop}%
\bibitem [{\citenamefont {Gan}\ \emph {et~al.}(2021)\citenamefont {Gan},
  \citenamefont {Liu}, \citenamefont {Sum},\ and\ \citenamefont
  {Hippalgaonkar}}]{Gan21v259}%
  \BibitemOpen
  \bibfield  {author} {\bibinfo {author} {\bibfnamefont {C.~K.}\ \bibnamefont
  {Gan}}, \bibinfo {author} {\bibfnamefont {Y.}~\bibnamefont {Liu}}, \bibinfo
  {author} {\bibfnamefont {T.~C.}\ \bibnamefont {Sum}},\ and\ \bibinfo {author}
  {\bibfnamefont {K.}~\bibnamefont {Hippalgaonkar}},\ }\bibfield  {title}
  {\bibinfo {title} {Efficacious symmetry-adapted atomic displacement method
  for lattice dynamical studies},\ }\href@noop {} {\bibfield  {journal}
  {\bibinfo  {journal} {Comput. Phys. Commun.}\ }\textbf {\bibinfo {volume}
  {259}},\ \bibinfo {pages} {107635} (\bibinfo {year} {2021})}\BibitemShut
  {NoStop}%
\bibitem [{\citenamefont {Gan}\ and\ \citenamefont {Ong}(2021)}]{Gan21v5}%
  \BibitemOpen
  \bibfield  {author} {\bibinfo {author} {\bibfnamefont {C.~K.}\ \bibnamefont
  {Gan}}\ and\ \bibinfo {author} {\bibfnamefont {Z.-Y.}\ \bibnamefont {Ong}},\
  }\bibfield  {title} {\bibinfo {title} {Complementary local-global approach
  for phonon mode connectivities},\ }\href@noop {} {\bibfield  {journal}
  {\bibinfo  {journal} {J. Phys. Commun.}\ }\textbf {\bibinfo {volume} {5}},\
  \bibinfo {pages} {015010} (\bibinfo {year} {2021})}\BibitemShut {NoStop}%
\bibitem [{\citenamefont {Gan}\ \emph {et~al.}(2010)\citenamefont {Gan},
  \citenamefont {Fan},\ and\ \citenamefont {Kuo}}]{Gan10v49}%
  \BibitemOpen
  \bibfield  {author} {\bibinfo {author} {\bibfnamefont {C.~K.}\ \bibnamefont
  {Gan}}, \bibinfo {author} {\bibfnamefont {X.~F.}\ \bibnamefont {Fan}},\ and\
  \bibinfo {author} {\bibfnamefont {J.-L.}\ \bibnamefont {Kuo}},\ }\bibfield
  {title} {\bibinfo {title} {Composition-temperature phase diagram of
  {B}e$_x${Z}n$_{1-x}${O} from first principles},\ }\href@noop {} {\bibfield
  {journal} {\bibinfo  {journal} {Comput. Mater. Sci.}\ }\textbf {\bibinfo
  {volume} {49}},\ \bibinfo {pages} {S29} (\bibinfo {year} {2010})}\BibitemShut
  {NoStop}%
\bibitem [{\citenamefont {Shiomi}\ \emph {et~al.}(2011)\citenamefont {Shiomi},
  \citenamefont {Esfarjani},\ and\ \citenamefont {Chen}}]{Shiomi11v84}%
  \BibitemOpen
  \bibfield  {author} {\bibinfo {author} {\bibfnamefont {J.}~\bibnamefont
  {Shiomi}}, \bibinfo {author} {\bibfnamefont {K.}~\bibnamefont {Esfarjani}},\
  and\ \bibinfo {author} {\bibfnamefont {G.}~\bibnamefont {Chen}},\ }\bibfield
  {title} {\bibinfo {title} {Thermal conductivity of half-{H}eusler compounds
  from first-principles calculations},\ }\href@noop {} {\bibfield  {journal}
  {\bibinfo  {journal} {Phys. Rev. B}\ }\textbf {\bibinfo {volume} {84}},\
  \bibinfo {pages} {104302} (\bibinfo {year} {2011})}\BibitemShut {NoStop}%
\end{thebibliography}%

\end{document}